\documentstyle[11pt,aaspp4]{article}

\def\be{\begin{equation}}
\def\ee{\end{equation}}
\def\eqref#1{(\ref{eqn:#1})}

\def\overdeng{\rho_g/{\overline \rho}_g}
\def\overdenm{\rho_m/{\overline \rho}_m}

\def\hmpc{{h^{-1}\;{\rm Mpc}}}

\def\kms{{\rm \;km\;s^{-1}}}
\def\hubunits{\kms\;{\rm Mpc}^{-1}}
\def\kunits{{h\;{\rm Mpc}^{-1}}}
\def\kunitskms{{\rm km}^{-1}\;s}

\def\lya{Ly-$\alpha$ }

\begin{document}

\title{LYMAN-ALPHA FOREST CONSTRAINTS ON THE MASS OF WARM DARK MATTER AND THE SHAPE OF THE LINEAR POWER SPECTRUM}

\author{
Vijay K. Narayanan\altaffilmark{1},
David N. Spergel\altaffilmark{1},
Romeel Dav\'e\altaffilmark{1,3}, and
Chung-Pei Ma\altaffilmark{2,4}
}
\altaffiltext{1}{Department of Astrophysical Sciences, Princeton University, Princeton, NJ 08544-1001; Email: vijay,dns,rad@astro.princeton.edu}

\altaffiltext{2}{Department of Physics and Astronomy, University of Pennsylvania, Philadelphia PA 19104; Email: cpma@physics.upenn.edu}

\altaffiltext{3}{Spitzer Fellow}

\altaffiltext{4}{Alfred P. Sloan Fellow}

\begin{abstract}
High resolution N-body simulations of cold dark matter (CDM) models
predict that  galaxies and clusters have cuspy halos with excessive 
substructure.
Observations reveal smooth halos with central density cores.  
One possible resolution of this conflict is that the dark matter is 
warm (WDM);  this will suppress the power spectrum on small scales.
The Lyman-alpha forest is a powerful probe of the linear power spectrum
on these scales.
We use collisionless N-body simulations to follow the evolution of 
structure in WDM models, and analyze artificial Lyman-alpha forest spectra 
extracted from them.
By requiring that there is enough small scale power in the linear power 
spectrum to reproduce the observed properties of the Lyman-alpha forest 
in quasar spectra, we derive a lower limit  to the mass 
of the WDM particle of 750 eV.  This limit is robust to reasonable  
uncertainties in our assumption about the temperature of the mean density gas
$(T_{0})$ at $z=3$.  
We argue that any model that suppresses the CDM linear theory power spectrum 
more severely than a 750 eV WDM particle cannot produce the Lyman-alpha forest.

\end{abstract}

\keywords{cosmology:theory, dark matter, methods:numerical, quasars: absorption lines}

\section{Introduction}

Cold dark matter (CDM) models of structure formation have been remarkably
successful in explaining a wide range of observations at both low and
high redshifts.
However, high resolution N-body simulations of the CDM model have revealed
two potential conflicts with observations.
First, the inner mass density profiles of simulated halos are cuspy, as 
opposed to the shallow profiles inferred from the rotation curves
of dwarfs and low surface brightness galaxies 
(\cite{moore94}; \cite{deblok97}; \cite{hernandez98}; 
but see van den Bosch et al. 1999),
and the observations of soft cores in galaxy clusters by gravitational 
lensing (\cite{tyson98}).
Second, group-sized halos in simulations contain a large number 
of low mass subhalos, greatly in excess of the observed number of satellite 
galaxies in the Local Group (\cite{klypin99}; \cite{moore99}).
A variety of mechanisms have been proposed recently to reduce the clustering 
of mass on small scales, while simultaneously retaining the large scale
successes of the CDM model.
These mechanisms include breaking the scale invariance of the power spectrum 
of primordial density fluctuations (\cite{kamionkowski99}), postulating 
different collapse histories for dark matter and baryons (\cite{navarro96}; 
\cite{bullock00}; \cite{binney00}), and modifying the nature of dark matter.
The last option includes dark matter models with a variety 
of properties --- self-interacting (\cite{spergel00}),
warm (\cite{sommerlarsen99})
repulsive (\cite{goodman00}), fluid (\cite{peebles00}), 
and fuzzy (\cite{hu00}).

In the warm dark matter (WDM) model, the linear power spectrum
is exponentially damped on scales smaller than the free-streaming length of
the warm particle,
$R_{f} = 0.2(\Omega_{{\rm W}}h^{2})^{1/3}(m_{{\rm W}}/{\rm keV})^{-4/3}\ {\rm Mpc}$, relative to the pure CDM model 
(\cite{bardeen86}).
Here, $\Omega_{{\rm W}}$ is the ratio of WDM density to the critical
density, $m_{{\rm W}}$ is the mass of the WDM particle, and 
$h\equiv H_0/100\hubunits$ is the Hubble parameter.
Non-linear gravitational evolution transfers power from large scales to
small scales, so the clustering of the highly nonlinear mass distribution 
is insensitive to the shape of the linear power spectrum below the non-linear
scale (\cite{lwp91}; \cite{bagla97}; \cite{white00}).
Conversely, \lya absorbers seen in the spectra of high redshift ($z\sim 3$) 
quasars arise from mass density fluctuations in the quasi-linear regime
(\cite{bi93}; \cite{hernquist96}; \cite{hui97}), so their properties 
remain sensitive to the linear power spectrum.
In this $Letter$, we set a lower limit on the mass of WDM particle 
by requiring that there be enough small scale power in the initial conditions 
to reproduce the observed properties of the \lya forest at $z=3$.

\section{Simulations}

We study the \lya forest in both CDM and WDM models with 
$\Omega_{m} = 0.4$, $\Omega_{\Lambda} = 0.6$, $h=0.65$,
$\Omega_{b}h^{2} = 0.02$, and $\sigma_{8m} = 0.95$, 
where $\Omega_{m}$, $\Omega_{\Lambda},$ and 
$\Omega_{b}$ are the contributions from total mass, vacuum energy, and baryons
to the total energy density of the universe,
and $\sigma_{8m}$ is the rms density fluctuation in 
$8 \hmpc$ spheres, chosen here to reproduce the
observed cluster abundance (\cite{white93}; \cite{eke96}).
Table 1 lists the parameters of all models investigated.
Our WDM models have 
$m_{{\rm W}} = 200, 500, 750,$ and $1000$ eV (corresponding to
$R_{f} = 0.95, 0.28, 0.16, \ {\rm and\ } 0.11 {\rm Mpc}$),
spanning the range of WDM masses required
to match the phase space density of dark matter cores in disk galaxies and 
dwarf spheroidal galaxies 
(\cite{hogan00} and references therein; \cite{sellwood00}) 
We also analyze a Broken Scale Invariance model (BSI, 
\cite{kamionkowski99}),
using an analytic fit to its power spectrum from White \& Croft (2000),
with a filtering scale $k_{0} = 2 \kunits$ required to fit 
the observed number density of low mass satellite galaxies in the Local Group.

\begin{table}
\caption{Model parameters.}
\bigskip
\begin{tabular}{llccc}
\tableline\tableline
Model & Power spectrum & $m_{\rm W}$ & $v_{\rm rms}(z_{i}=49)$ & $T_{0} (z=3)$\\
& & (eV) & ($\kms$) & (K) \\
\tableline
CDM  & CDM	      & -- & 0 & 6000 \\
WDM1000 & WDM & 1000 & 2.3 & 6000 \\
WDM750  & WDM & 750 & 3.3 & 6000 \\
WDM500  & WDM & 500 & 5.7 & 6000 \\
WDM200  & WDM & 200 & 19.2 & 6000 \\
CDM200  & WDM\tablenotemark{*} & 200 &  0  & 6000 \\
WDM750T025K & WDM & 750 & 3.3 & 25000 \\
BSIK02  & CDM,BSI\tablenotemark{\dagger} & -- & 0 & 6000 \\
\tableline
\end{tabular}
\tablenotetext{*}{This model has the same linear power spectrum as WDM200.}
\tablenotetext{\dagger}{We use an analytic fit to the power spectrum of 
the Broken Scale Invariance model with $k_{0}= 2\kunits$, given by White \& Croft 2000.
}
\label{table:models}
\end{table}
We calculate the linear power spectrum of the mass density field for all
the models using the full Boltzmann transport code of Ma \&
Bertschinger (1995).  
We assume the usual three massless neutrino species with a present-day
temperature $T_{0,\nu}=1.947$ K, and treat the WDM component
as a fourth (massive) species with 
$T_{0,W} = T_{0,\nu} (\Omega_W\,h^2)^{1/3} (93\,{\rm eV}/m_W)^{1/3}$.
Fifty Legendre moments are used to follow the evolution of the WDM
phase space. 
We compared our WDM power spectra with the fitting function of
Bardeen et al. (1986) and find that their formula provides a
reasonable approximation for $k < 5 \kunits$  [if a
baryon-corrected shape parameter $\Gamma=\Omega_m h
\exp(-\Omega_b\,(1+\Omega_m^{-1}))$ is used], but it overestimates
the power spectrum by up to an order of magnitude at higher $k$.

We employ a particle-mesh (PM) N-body code that is described in detail in 
Steed et al. (in preparation).
This code computes forces using a staggered mesh
(\cite{melott86}; \cite{park90}), and integrates the equations of motion 
using the leapfrog scheme described in Quinn et al. (1997).
Our periodic simulation volume has $L=25 h^{-1}$Mpc, with
$N_p=256^{3}$ particles and an $N_m=512^{3}$ force mesh.
We assign initial displacements and velocities to the particles using the
Zeldovich approximation.
In the WDM models, we also add a randomly oriented streaming 
velocity component that is drawn from a Fermi-Dirac distribution with 
an rms velocity,
$
v_{\rm rms} = 48.7 \left(\frac{h}{0.65}\right )^{\frac{2}{3}}\left(\frac{m_{\rm W}}{100\ {\rm eV}}\right )^{-\frac{4}{3}}\left(\frac{\Omega_{\rm W}}{0.4}\right )^{\frac{1}{3}} \left(\frac{1+z_{\rm i}}{50}\right) \kms.
$
We evolve from redshift $z=49\rightarrow 3$ 
in 24 equal steps of the expansion scale factor.


Figure 1 shows the linear and non-linear power spectrum of the mass
density field at $z=3$ in different models.
The differences between different models are significantly smaller in 
the non-linear power spectra, compared to the differences in their
linear theory power spectra, because
non-linear gravitational evolution has regenerated power on 
small scales.
Nevertheless, power transfer 
is not entirely successful in erasing the differences between the power spectra
of different models.
Thus, at $z=3$, the WDM200 model has almost an order of magnitude less 
power on small scales compared to the CDM model, while the models
WDM500, WDM750, and BSIK02 are all deficient by $\sim 0.3$~dex.
We also ran these models with a $50 \hmpc$ box using the same $N_p$ and $N_m$,
and found similar results on overlapping length scales.

The model CDM200 has the same linear power spectrum as WDM200,
but $v_{\rm rms} = 0$.
This model gives virtually identical results to the WDM200 model,
showing that the velocity dispersion of WDM particles has a negligible effect 
on the \lya forest at $z=3$. 
Hence, we do not show the results for this model in the figures.

\begin{figure}
\centerline{
\epsfxsize=\hsize
\epsfbox[18 300 592 650]{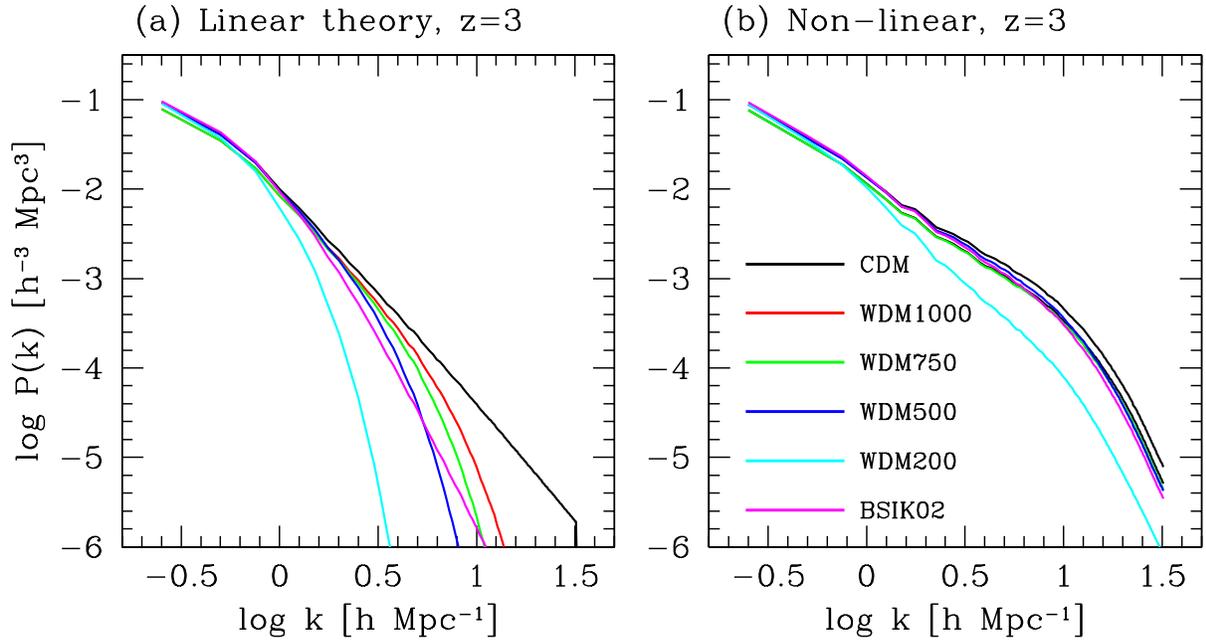}
}
\caption{ Power spectra of the mass density fields in different models.
(a) Linear mass power spectra at $z=3$ from the Boltzmann code.
(b) Non-linear power spectra at $z=3$ from simulation output,
computed by cloud-in-cell binning onto a $256^{3}$ grid, taking its Fourier 
transform, and squaring the amplitudes of its Fourier components.
Note that WDM750T025K has the same {\it mass} power spectrum as
WDM750.
\label{fig:pspz3}
}
\end{figure}

\section{Properties of the \lya forest}

The \lya forest arises from a continuous, fluctuating, low-density
$[-1 \la {\rm log (\overdeng)} \la 1]$ gaseous medium whose temperature 
is determined by the balance between photoionization heating by the UV 
background and adiabatic cooling due to Hubble expansion 
(see e.g., \cite{bi93}; \cite{cen94}; \cite{zhang95}; \cite{hernquist96}; 
\cite{hui97}).
Photoionization equilibrium leads to a power-law relation between 
temperature and gas density, 
of the form $T = T_{0}(\overdeng)^{\alpha}$,
with $ 4000{\rm \;K} \le T_{0} \le 10000{\rm\;K}$, and 
$0.3 \le \alpha \le 0.6$ (\cite{hui97}).
The optical depth to \lya photons ($\tau_{{\rm Ly}\alpha}$) is proportional 
to \ion{H}{1} gas density, which in turn is proportional to 
$T^{-0.7}$ for gas in photoionization equilibrium near 
$10^{4}{\rm \; K}$.
Pressure gradients in this low-density, fairly cool ($T \sim 10^{4}$ K) 
gas are small, so gas traces dark matter (\cite{croft98}).
Hence, the gas density-temperature relation can be translated into a relation
between mass density $(\rho_{m})$ and $\tau_{{\rm Ly}\alpha}$
(the ``Fluctuating Gunn-Peterson Approximation", 
\cite{rauch97}; \cite{croft98}; \cite{weinberg98}).
Thus, the transmitted flux of \lya photons 
$[{\rm F} = \exp(-\tau_{{\rm Ly}\alpha})]$ in a quasar spectrum is a map 
of the mass density fluctuations along the line of sight.

We extract artificial \lya absorption spectra along 400 random lines of 
sight through each simulation, using the TIPSY package 
(\cite{tipsy95}; see http://www-hpcc.astro.washington.edu/tools/TIPSY).
TIPSY calculates the local mass density $(\rho_{m})$ at the position of each
dark matter particle using a cubic spline smoothing kernel
(Hernquist \& Katz 1989) enclosing 32 neighbors, and assigns it a
temperature $T = T_0(\overdenm)^{0.6}$.
Spectra are then computed from this particle distribution using
the algorithm described by Hernquist et al.\ (1996).
We sample the spectra with resolution $\Delta v = 6.4 \kms$, roughly
corresponding to that of the Keck HIRES spectra to which we 
will compare our model predictions.
We ``fit a continuum'' to each spectral segment, $3200 \kms$ in length,
by rescaling all flux values so that the highest pixel has transmission
of unity (\cite{dave99}).  
Fitting a continuum makes low-power models appear more like high power 
models, so it is an important step for conservatively discriminating against 
models with suppressed power.

The optical depth to \lya photons depends on a number of uncertain
parameters including $\Omega_{b}$, $h$, $T_{0}$, and the intensity of the 
ionizing background.
However, the cumulative effect of all these parameters is to scale
the normalization factor ($A$) relating $\tau_{{\rm Ly}\alpha}$
to $\overdenm$, as long as collisional ionization is unimportant 
(\cite{weinberg98a}).
We fix $A$ in each model so that the mean transmitted flux at $z=3$,
${\overline {\rm F}} = \left< \exp(-\tau_{{\rm Ly}\alpha}) \right>$, is
0.684, as measured by McDonald et al. (1999) from Keck HIRES spectra of 
eight quasars.

Figure 2 shows artificial \lya absorption spectra at $z=3$ extracted along 
four random lines of sight, in three different models:
CDM, WDM500, and WDM750.
The gas responsible for \lya absorption is in larger and more diffuse 
structures in WDM models compared to the CDM model.
Hence, the WDM spectra are smoother and have less small scale power 
as compared to the CDM spectra.
The fluctuations in the absorbing gas are also weaker in the 
WDM models, resulting in
absorption lines that are broader and 
shallower than corresponding lines in the CDM artificial spectra.


\begin{figure}
\centerline{
\epsfxsize=5.5truein
\epsfbox[70 80 550 450]{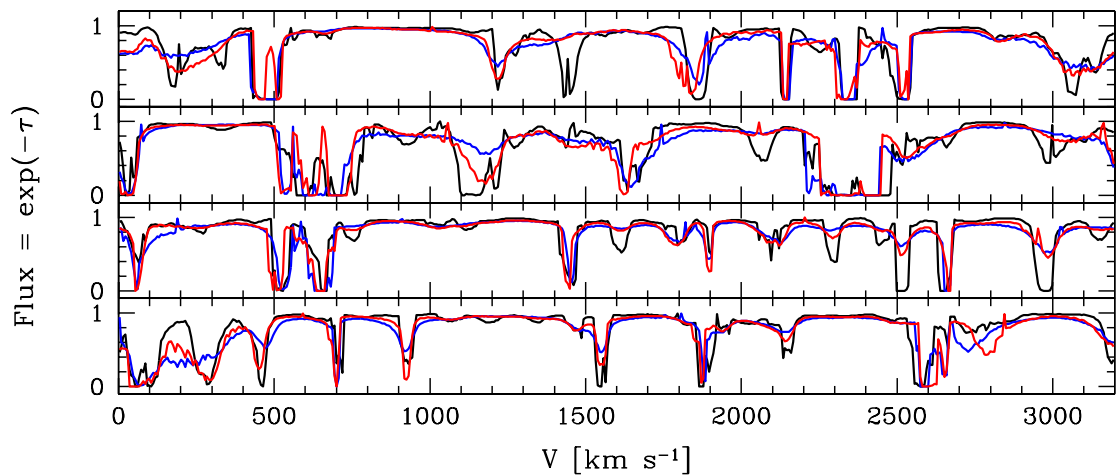}
}
\caption{Examples of artificial spectra at $z=3$ along four random lines of 
sight in the simulations. All simulations have identical phases for the Fourier
components of the initial density fields.
In each panel, black line shows results for CDM, blue line shows WDM500,
and red line shows WDM750.
\label{fig:spectra}
}
\end{figure}

Figure 3 shows the one dimensional power spectrum of the transmitted
flux, $P_{F}(k)$, computed using 400 artificial spectra for each model.
The solid points show $P_{F}(k)$ measured by McDonald et al. 
(1999).
The shape and amplitude of $P_{F}(k)$ of the CDM model 
matches the observations very well, 
in the entire range $-2.5 \la \log [k (\kunitskms)] \la -1$.
However, the WDM models predict a wrong shape of $P_{F}(k)$.
Thus, although the $P_{F}(k)$ of WDM models match the 
observations on large scales,  $-2.5 \la \log [k (\kunitskms)] \la -2$,
they have too little power on smaller scales.
This confirms the visual impression from the spectra in Fig. 2, and is
also evident in the flux decrement correlation function (not shown).
A higher value of $T_{0}$ further suppresses $P_{F}(k)$ on smaller scales, 
worsening the discrepancy with observations.
Note that lowering $\Omega_m$ to make up the power suppressed by WDM is not
feasible; a lower $\Omega_m$ raises the power spectrum on all
scales at $z=3$, and thus would be inconsistent with the observed $P_{F}(k)$ 
on large scales.


\begin{figure}
\centerline{
\epsfxsize=\hsize
\epsfbox[18 144 592 718]{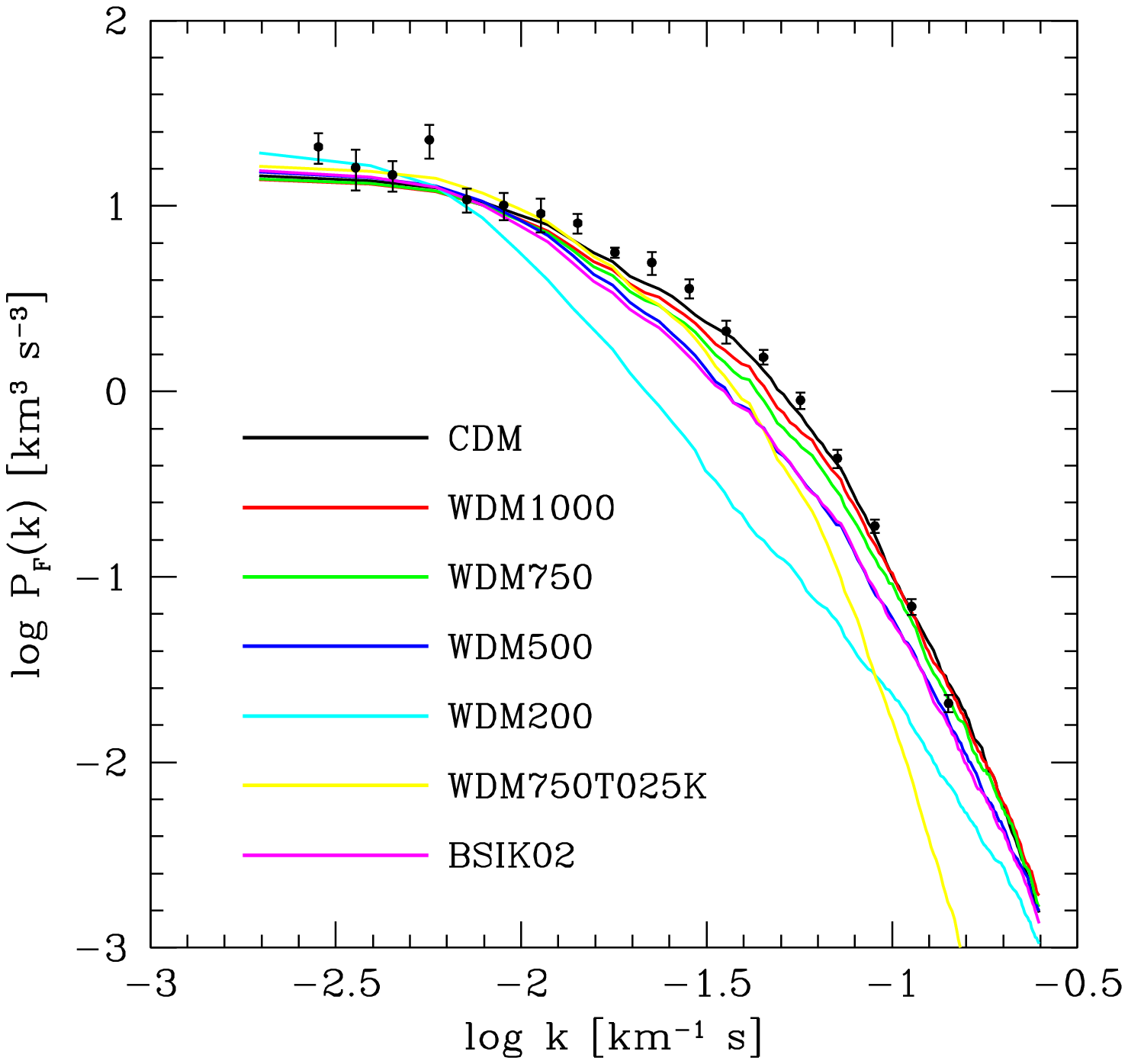}
}
\caption{ 
Power spectrum of the transmitted flux at $z=3$ from 400 artificial spectra
for each model.
Solid points with error bars show the flux power spectrum at $z\approx 3$ 
measured by McDonald et al. (1999) from eight Keck HIRES quasar spectra.
\label{fig:fpk1d}
}
\end{figure}

Figure 4 shows the probability distribution function (PDF) of the transmitted
flux (\cite{jenkins91}; \cite{jordi96}; \cite{rauch97}), computed using 
400 artificial spectra for each model.
The solid points show the flux PDF measured by McDonald et al. (1999).
The flux PDF of the CDM model matches observations remarkably well over
the entire range of flux values.
As the mass of WDM decreases, there is less small scale power in the 
mass distribution, the mass fluctuations are weaker, and
the gas responsible for \lya absorption is in more diffuse structures.
Hence there are fewer pixels where the transmission approaches unity,
and more pixels at intermediate fluxes.
In the flux range $0.6 < {\rm F} < 0.8$, the flux PDF for all WDM models with
$m_{\rm W} \le 750\ {\rm eV}$ differs from observations
by $\ga 3\sigma$.
This mismatch further confirms that the mass fluctuation amplitude of these
WDM models is too low at small scales when the fluctuations are normalized
to match $P_{F}(k)$ on large scales.
Even the model WDM750T025K, with a higher temperature for the 
mean-density gas ($T_{0} = 25,000\;{\rm K}$),
does not match the observations as well as the CDM model.
On the other hand, while WDM1000 does not match the observations
very well, a higher value of $T_{0}$ can bring its flux PDF to better
agreement with observations.
Since recent determinations of $T_{0}$ prefer values around 
$20,000\;{\rm K}$ (e.g., \cite{schaye99}), we conclude that the observed flux
PDF cannot be reproduced by WDM models with $m_{\rm W}\la 750$ eV.
The BSIK02 model is also inconsistent with observations, being similar
to WDM500.


\begin{figure}
\centerline{
\epsfxsize=\hsize
\epsfbox[18 144 592 718]{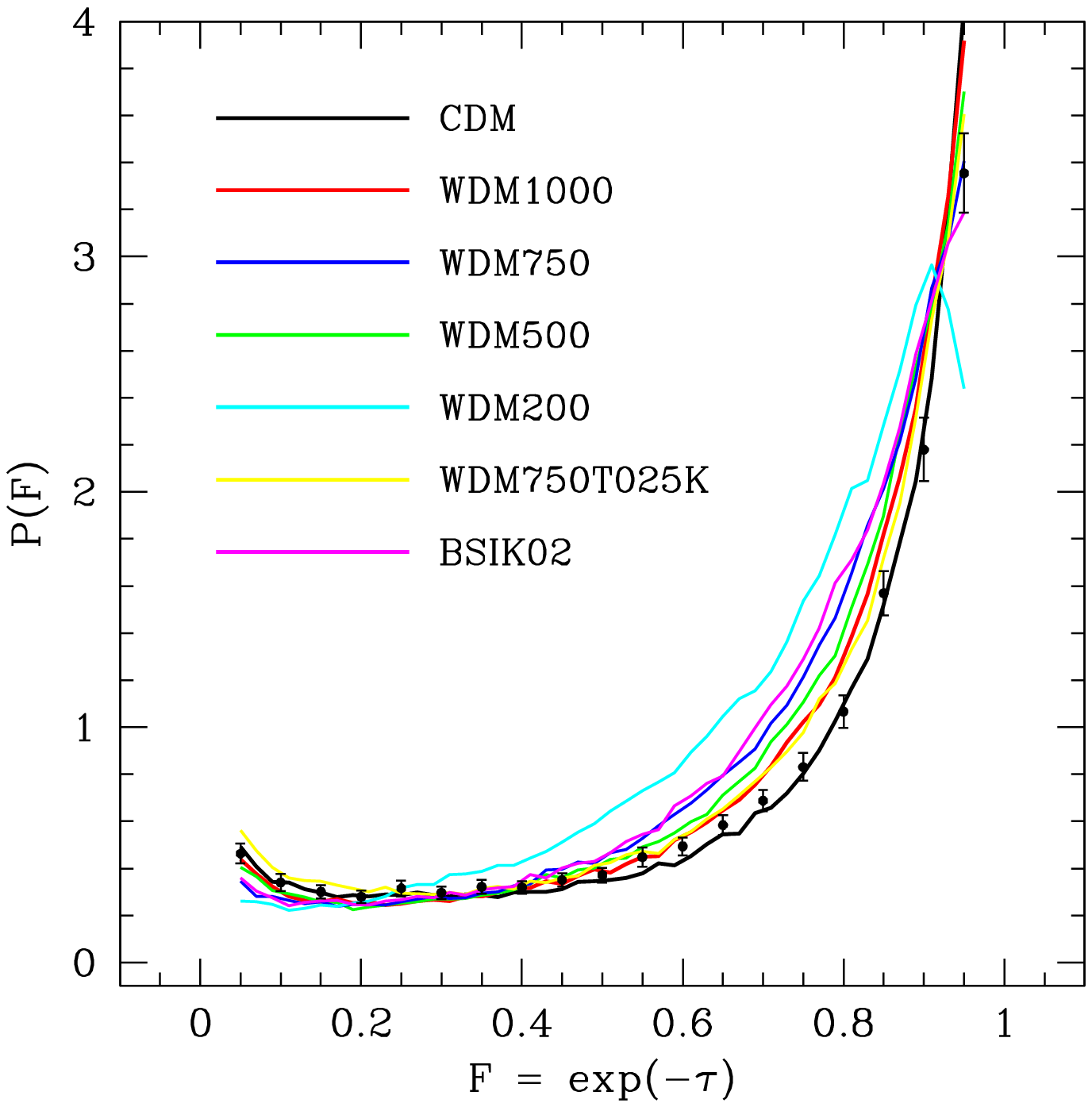}
}
\caption{ Probability distribution functions (PDF) of the transmitted flux at 
$z=3$, computed using 400 artificial spectra for each model.
Solid points with error bars show the flux PDF at $z=3$ measured by
McDonald et al. (1999) using Keck HIRES spectra of eight quasars.
\label{fig:fpdf}
}
\end{figure}

\section{Conclusions}

For a given cosmology, the properties of the \lya forest depend on
three factors: (a) the linear theory power spectrum, taken from a
Boltzmann code; (b) power transfer from large to small
scales during non-linear gravitational evolution, followed using N-body
simulations; and (c) the equation of state of the intergalactic medium,
governed by photoionization equilibrium.
Here, we have used the observed \lya forest at $z=3$ to constrain the shape 
of the linear power spectrum, and consequently $m_{\rm W}$.
Our principal conclusions are as follows:
\newline
(1) $m_{\rm W}\ga 750$ eV is required to
match $P_{F}(k)$ and flux PDF of artificial 
spectra extracted from simulations of WDM models with Keck HIRES
observations of McDonald et al.\ (1999), for any reasonable $T_0$.
This limit is similar to that obtained from the maximum observed phase space 
density of dwarf spheroidal galaxies (Dalcanton \& Hogan 2000).
High resolution N-body simulations of WDM models with $m_{\rm W} \sim 1$ keV 
predict the observed circular velocity function of satellite galaxies 
in the Local Group, but produce cuspy halos (\cite{colin00}).
\newline
(2) \lya forest properties depend on the shape of the linear
power spectrum, and are insensitive to streaming velocities of WDM particles.
Even for a 200~eV WDM particle, 
$v_{\rm rms} = 1.5 \kms$ at $z=3$, an order of magnitude smaller than the
thermal velocity of \ion{H}{1} gas at $T \sim 10^{4}$~K.
Hence, any mechanism that suppresses the CDM linear power spectrum 
more severely than a 750~eV WDM particle will be inconsistent with
\lya forest observations.
\newline
(3) The  BSI model proposed by Kamionkowski \& Liddle (1999), 
with $k_{0} = 2\kunits$ chosen to match the abundance of dwarf galaxies, 
does not reproduce the observed properties of the 
\lya forest at $z=3$.  We note that this is the extreme BSI model considered 
by  White \& Croft (2000), hence our conclusion is not in conflict with theirs.

\acknowledgments
We thank David Weinberg and Michael Strauss for helpful comments on the 
manuscript.
DNS is partially supported by the MAP/MIDEX program, and the
NASA ATP grant NAG5-7154.
CPM is supported by a Cottrell Scholars
Award from the Research Corporation, and NSF grant AST 9973461.


\begin{thebibliography}{wdm00}

\bibitem[Bagla \& Padmanabhan 1997]{bagla97}
Bagla, J. S., \& Padmanabhan, T. 1997, \mnras, 286, 1023

\bibitem[Bardeen et al.\ 1986]{bardeen86}
Bardeen, J., Bond, J. R., Kaiser, N., \& Szalay, A. 1986, \apj, 304, 15

\bibitem[Bi\ 1993]{bi93}
Bi, H., 1993, \apj, 405, 479

\bibitem[Bi \& Davidsen\ 1997]{bi97}
Bi, H., \& Davidsen, A. F.,\ 1997, \apj, 479, 523

\bibitem[Binney, Gerhard, \& Silk\ 2000]{binney00}
Binney, J. J., Gerhard, O., \& Silk, J., 2000, \mnras, submitted (astro-ph/0003199)

\bibitem[Bullock, Kravtsov, \& Weinberg\ 2000]{bullock00}
Bullock, J. S., Kravtsov, A. V., \& Weinberg, D. H, 2000, \apj, submitted (astro-ph/0002214)

\bibitem[Colin, Avila-Reese, \& Valenzuela\ 2000]{colin00}
Colin, P., Avila-Reese, V., \& Valenzuela, O., 2000, \apj, submitted (astro-ph/0004115)

\bibitem[Cen et al.\ 1994]{cen94}
Cen, R., Miralda-Escude, J., Ostriker, J. P., \& Rauch, M., 1994, \apj, 437, L9


\bibitem[Croft et al.\ 1998]{croft98}
Croft, R. A. C., Weinberg, D. H., Katz, N., \& Hernquist, L. 1998,
\apj, 495, 44

\bibitem[Dalcanton \& Hogan\ 2000]{dalcanton00}
Dalanton, J. J,. \& Hogan, C. J., 2000, preprint (astro-ph/0004381)

\bibitem[Dav\'e et al.\ 1999]{dave99}
Dav\'e, R., Hernquist, L., Katz, N., \& Weinberg, D. H., 1999, \apj, 511, 521

\bibitem[de Blok \& McGaugh\ 1997]{deblok97}
de Blok, W. J. G., \& McGaugh, S. S.,\ 1997, \mnras, 290, 533

\bibitem[Eke, Cole, \& Frenk 1996]{eke96} 
Eke, V. R., Cole, S., Frenk, C. S. 1996, \mnras, 282, 263

\bibitem[Flores \& Primack\ 1994]{flores94}
Flores, R., \& Primack, J. R., 1994, \apj, 427, L1

\bibitem[Goodman 2000]{goodman00}
Goodman, J., 2000, preprint (astro-ph/0003018)

\bibitem[Hernandez \& Gilmore\ 1998]{hernandez98}
Hernandez, X., \& Gilmore, G., 1998, \mnras, 294, 595

\bibitem[Hernquist \& Katz 1989]{hernquist89}
Hernquist, L., \& Katz, N. 1989, \apjs, 70, 419

\bibitem[Hernquist et al.\ 1996]{hernquist96}
Hernquist L., Katz, N., Weinberg, D. H., \& Miralda-Escud\'e, J. 1996, \apj, 457, L5

\bibitem[Hogan \& Dalcanton\ 2000]{hogan00}
Hogan, C. J., \& Dalanton, J. J., 2000, preprint (astro-ph/0002330)

\bibitem[Hu, Barkana, \& Gruzinov\ 2000]{hu00}
Hu, W., Barkana, R., \& Gruzinov, A.,\ 2000, Physics Review Letters, submitted (astro-ph/0004151)

\bibitem[Hui \& Gnedin\ 1997]{hui97}
Hui, L., \& Gnedin, N. Y., 1997, \mnras, 292, 27

\bibitem[Jenkins \& Ostriker \ 1991]{jenkins91}
Jenkins, E. B., \& Ostriker, J. P.,\ 1991, \apj, 376, 33

\bibitem[Kamionkowski \& Liddle 1999]{kamionkowski99}
Kamionkowski, M.,  \& Liddle, A. R., 1999, preprint (astro-ph/9911103)

\bibitem[Katz \& Quinn\ 1995]{tipsy95}
Katz, N., \& Quinn, T., 1995, TIPSY manual

\bibitem[Klypin et al.\ 1999]{klypin99}
Klypin, A., Kravtsov, A. V., Valenzuela, O., \& Prada, F., 1999, \apj, 522, 82

\bibitem[Little, Weinberg, \& Park 1991]{lwp91} 
Little, B., Weinberg, D. H, \& Park, C. 1991, \mnras, 253, 295

\bibitem[Ma \& Bertschinger\ 1995]{ma95}
Ma, C. P., \& Bertschinger, E. B., 1995, \apj, 455, 7

\bibitem[McDonald et al.\ 1999]{mcdonald99}
McDonald, P., Miralda-Escud\'e, J., Rauch, M., Sargent, W. L. W., Barlow, T. A., Cen, R., \& Ostriker, J. P., \apj, submitted (astro-ph/9911196)

\bibitem[Miralda-Escude et al.\ 1996]{jordi96}
Mirald-Escude, J., Cen, R., Ostriker, J. P., \& Rauch, M., 1996, \apj, 471, 582

\bibitem[Melott 1986]{melott86}
Melott, A. L., 1986, Physics Review Letters, 56, 1992

\bibitem[Moore 1994]{moore94}
Moore, B., 1994, Nature, 370, 629

\bibitem[Moore et al. 1999]{moore99}
Moore, B., Ghigna, S., Governato, F., Lake, G., Quinn, T., Stadel, J., \& Tozzi, P., 1999, \apj, 524, L19

\bibitem[Moore et al.\ 2000]{moore00}
Moore, B., Gelato, S., Jenkins, A., Pearce, F., \& Quilis, V., 2000, \mnras, submitted (astro-ph/0002308)

\bibitem[Navarro, Eke, \& Frenk 1996]{navarro96}
Navarro, J. F., Eke, V. R., \& Frenk, C. S., 1996, \mnras, 283, L72

\bibitem[Park\ 1990]{park90}
Park, C., Ph.D Thesis, Princeton University.

\bibitem[Peebles 2000]{peebles00}
Peebles, P. J. E., 2000, preprint (astro-ph/0002495)

\bibitem[Quinn et al.\ 1997]{quinn97}
Quinn, T., Katz, N., Stadel, J., \& Lake, G., 1997, preprint (astro-ph/9710043)

\bibitem[Rauch et al.\ 1997]{rauch97}
Rauch, M., et al., 1997, \apj, 489, 7

\bibitem[Schaye et al.\ 1999]{schaye99}
Schaye, J., Theuns, T., Rauch, M., Efstathiou, G., \& Sargent, W. L. W., 1999, \mnras, submitted (astro-ph/9912432)

\bibitem[Sellwood 2000]{sellwood00}
Sellwood, J. A., 2000, \apj, submitted (astro-ph/0004352)


\bibitem[Sommer-Larsen \& Dolgov\ 1999]{sommerlarsen99}
Sommer-Larsen, J., \& Dolgov, A.,\ 1999, \apj, submitted (astro-ph/9912166)

\bibitem[Spergel \& Steinhardt\ 2000]{spergel00}
Spergel, D. N., \& Steinhardt, P. J., 2000, Physics Review Letters, in press

\bibitem[Tyson, Kochanski, \& Dell\'Antonio\ 1998]{tyson98}
Tyson, J. A., Kochanski, G. P., \& Dell\'Antonio, I. P., 1999, \apj, 498, L107

\bibitem[van den Bosch et al.\ 1999]{vandenbosch99}
van den Bosch, F., Robertson, B. E., Dalcanton, J. J., \& de Blok, W. J. G., 1999, \aj, submitted (astro-ph/9911372)

\bibitem[Weinberg, Katz, \& Hernquist 1998]{weinberg98}
Weinberg, D. H., Katz, N., \& Hernquist, L. 1998, in ASP Conference Series 148, Origins, eds. C. E. Woodward, J. M. Shull, \& H. Thronson,
(ASP: San Francisco), 21, astro-ph/9708213

\bibitem[Weinberg et al.\ 1998]{weinberg98a}
Weinberg, D. H., et al. 1998, in Proceedings of the MPA/ESO Conference
"Evolution of Large Scale Structure: From Recombination to Garching"

\bibitem[White \& Croft\ 2000]{white00}
White, M., \& Croft, R. A. C., 2000, \apj, submitted (astro-ph/0001247)

\bibitem[White, Efstathiou, \& Frenk\ 1993]{white93}
White, S. D. M., Efstathiou, G. P., \& Frenk, C. S. 1993, \mnras, 262, 1023

\bibitem[Zhang, Anninos, \& Norman\ 1995]{zhang95}
Zhang, Y., Anninos, P., \& Norman, M. L., 1995, \apj, 453, L57

\end{thebibliography}
\end{document}